# A radio ridge connecting two galaxy clusters in a filament of the cosmic web

**Authors:** F. Govoni[1*], E. Orrù[2], A. Bonafede[3,4,5], M. Iacobelli[2], R. Paladino[3], F. Vazza[3,4,5], M. Murgia[1], V. Vacca[1], G. Giovannini[3,4], L. Feretti[3], F. Loi[1,4], G. Bernardi[3,6,7], C. Ferrari[8], R.F. Pizzo[2], C. Gheller[9], S. Manti[10], M. Brüggen[5], G. Brunetti[3], R. Cassano[3], F. de Gasperin[5,11], T.A. Enßlin[12,13], M. Hoeft[14], C. Horellou[15], H. Junklewitz[16], H.J.A. Röttgering[11], A.M.M. Scaife[17], T.W. Shimwell[2,11], R.J. van Weeren[11], M. Wise[2,18].

**Affiliations:**

[1] Istituto Nazionale di Astrofisica - Osservatorio Astronomico di Cagliari Via della Scienza 5, I09047 Selargius, Italy.

[2] ASTRON, the Netherlands Institute for Radio Astronomy, Postbus 2, 7990 AA, Dwingeloo, The Netherlands.

[3] Istituto Nazionale di Astrofisica - Istituto di Radioastronomia, Bologna Via Gobetti 101, I40129 Bologna, Italy

[4] Dip. di Fisica e Astronomia, Università degli Studi di Bologna, Via Gobetti 93/2, I40129 Bologna, Italy

[5] Hamburger Sternwarte, Universität Hamburg, Gojenbergsweg 112, 21029, Hamburg, Germany

[6] Department of Physics and Electronics, Rhodes University, Grahamstown, South Africa

[7] Square Kilometre Array South Africa, 3rd Floor, The Park, Park Road, Pinelands, 7405, South Africa

[8] Université Côte d'Azur, Observatoire de la Côte d'Azur, Centre national de la recherche scientifique, Laboratoire Lagrange, Blvd de l'Observatoire, CS 34229, 06304 Nice cedex 4, France

[9] Swiss Plasma Center, Ecole Polytechnique Fédérale de Lausanne, 1015 Ecublens, Lausanne, Switzerland

[10] Scuola Normale Superiore, Piazza dei Cavalieri 7, I-56126 Pisa, Italy

[11] Leiden University, Rapenburg 70, 2311 EZ Leiden, the Netherlands

[12] Max Planck Institut für Astrophysik, Karl-Schwarzschild-Str.1, D-85740 Garching, Germany

[13] Ludwig-Maximilians-Universität München, Geschwister-Scholl-Platz 1, D-80539, München, Germany

[14] Thüringer Landessternwarte, Sternwarte 5, 07778 Tautenburg, Germany

[15] Chalmers University of Technology, Dept. of Space, Earth and Environment, Onsala Space Observatory, SE-439 92 Onsala, Sweden

[16] Argelander-Institut für Astronomie, Auf dem Hügel 71 D-53121 Bonn, Germany

[17] Jodrell Bank Centre for Astrophysics, School of Physics and Astronomy, The University of Manchester, Oxford Road, Manchester M13 9PL, UK

[18] SRON Netherlands Institute for Space Research Sorbonnelaan 2, 3584 CA Utrecht, The Netherlands

*Correspondence to: federica.govoni@inaf.it

**Abstract:** Galaxy clusters are the most massive gravitationally bound structures in the Universe. They grow by accreting smaller structures in a merging process that produces shocks and turbulence in the intra-cluster gas. We observed a ridge of radio emission connecting the merging galaxy clusters Abell 0399 and Abell 0401 with the Low Frequency Array (LOFAR) at 140 MHz. This emission requires a population of relativistic electrons and a magnetic field located in a filament between the two galaxy clusters. We performed simulations to show that a volume-filling distribution of weak shocks may re-accelerate a pre-existing population of relativistic particles, producing emission at radio wavelengths that illuminates the magnetic ridge.

**One Sentence Summary:** Discovery of radio emission from a cosmic web filament located between two massive galaxy clusters.

**Main Text:** The matter distribution of the Universe is not uniform, but forms a cosmic web, with a structure of filaments and galaxy clusters surrounding large voids. Galaxy clusters form at the intersections of the cosmic web filaments and grow by accreting substructures in a merging process, which converts most of the infall kinetic energy into thermal gas energy. A residual fraction of non-thermalised energy is expected to manifest itself in the form of turbulent gas motions, magnetic fields, and relativistic particles. Extended radio sources called radio halos and radio relics are found at the center and the periphery of galaxy clusters, respectively, visible through their emission of synchrotron radiation. Magnetic fields and relativistic particles in the large-scale structure of the Universe can be inferred from observations of these sources.

Observations show that magnetic fields are ubiquitous in galaxy clusters (*1*), while radio halos and relics are most common in merging clusters, suggesting that cluster mergers provide the acceleration of relativistic particles necessary for synchrotron emission (*2*). Collisions between nearly equal mass clusters dissipate large amounts of energy within a few Gyr. The merging galaxy clusters Abell 0399 and Abell 0401 are separated by a projected distance of 3 Mpc and host a double radio halo (*3*). The presence of radio halos at the centers of both Abell 0399 and Abell 0401 was unexpected because radio halos are not common sources and X-ray (*4 – 7*) and optical data (*8*) suggest that the two systems are still in the initial phase of a merger, before the bulk of kinetic energy of the collision has been dissipated. X-ray data show a hot (temperature $T \simeq 6 - 7$ keV) and nearly isothermal filament of plasma in the region between the two clusters (*7*). There may be a weak shock (Mach number $M < 2$) in the outer part of the filament, at about $650 - 810$ kiloparsecs (kpc) from the collision axis (equivalent to an angular offset of $8 - 10$ arcminutes). Observations with the Planck spacecraft (*9, 10*) detected a signal due to the Sunyaev Zeldovich (SZ) effect, revealing a large-scale filament of gas connecting the two systems. The SZ effect is a spectral distortion caused by inverse Compton scattering of the cosmic microwave background radiation by hot electrons ($T \sim$ keV), sensitive to the total thermal energy of the intervening medium.

We observed the region between Abell 0399 and Abell 0401 at radio wavelengths to investigate whether relativistic particles and magnetic fields exist on cosmic scales larger than those of galaxy clusters. Using the Low Frequency Array (LOFAR) at a central frequency $v$=140 MHz (corresponding to a wavelength λ=2.14 meters), we detect a filament of diffuse synchrotron emission connecting the two galaxy clusters.

Figure 1 shows our LOFAR observation of a diffuse radio ridge encompassing Abell 0399 - Abell 0401 at a resolution of 80″. Our analysis of LOFAR images at higher resolution (Figs. S3 and S4) shows that no extended discrete radio sources are present between the two galaxy clusters (*11*). The ridge encompassing Abell 0399 – Abell 0401 is not due to the blending of discrete sources, but is associated with the cosmic web filament connecting the two clusters (*11*).

To measure the physical parameters of the ridge encompassing Abell 0399 – Abell 0401, we masked the regions occupied by the radio halos and other radio sources not connected to the ridge emission (*11*). Fig. 2A shows the radio ridge after the masking. It extends between the two cluster cores with a sky projected length of 3 Mpc. We extracted the brightness profile of the ridge (Fig. 2B) by computing the average brightness in strips of length 3 Mpc and width 0.108 Mpc (one beam width). We modelled the brightness profile using a Gaussian model characterized by three free parameters: the peak brightness, peak position, and full width at half maximum (FWHM). The ridge emission peaks at about 3.7 milliJansky (mJy) per beam with a FWHM of 1.3 Mpc. The ridge is offset to the north-west by 0.16 Mpc from the line connecting the cores of Abell 0399 and Abell 0401.

The flux density of the ridge was calculated by integrating the surface brightness over an area of 3×1.3 Mpc$^2$ after excluding the masked regions. The average surface brightness at 140 MHz is $<I>_{140MHz}$=2.75±0.08 mJy beam$^{-1}$ (or 0.38 μJy arcsec$^{-2}$). Taking into account the masked regions, the effective number of instrument beams covering the ridge emission, $N_{eff}$=160, is fewer than the 299 beams contained by this area. Assuming that the ridge is present everywhere, even in the masked regions, we extrapolate the total flux density $S_{140MHz} = <I>_{140MHz} \times 299$=822±24 (±123) mJy. The uncertainties are the 1σ root mean square (rms) statistical uncertainty and a 15% systematic uncertainty (indicated in parentheses), to account for the uncertain calibration of the LOFAR flux scale (*12*). This flux density corresponds to a radio power $L_{140MHz}$=1.0 × 10$^{25}$ W Hz$^{-1}$. By assuming the ridge occupies a cylindrical volume we estimate a mean radio emissivity $<J>_{140MHz}$=8.6×10$^{-43}$ erg s$^{-1}$Hz$^{-1}$cm$^{-3}$.

It is not possible to reliably determine the spectral index of the ridge emission, because the available data at 1.4 GHz (*3*) were obtained on different baselines. However, by adopting the spectral index α = 1.3 (where $S_\nu \propto \nu^{-\alpha}$) typically found in radio halos (*13*), the mean radio emissivity extrapolated to 1.4 GHz would be $<J>_{1.4GHz} \simeq$ 4.3 × 10$^{-44}$ erg s$^{-1}$ Hz$^{-1}$cm$^{-3}$. We compare this value with published distributions of emissivities at 1.4 GHz for candidate large scale filaments (*14*), and with radio halos observed at the center of galaxy clusters (*13, 15*). The histogram of the emissivity of the candidate filaments and of the radio halos (Fig. S2) has two distinct populations with only a partial overlap. The filament in Abell 0399 – Abell 0401 is located in the weakest tail of the emissivity distribution of the candidate filaments and is almost two orders of magnitude lower than the typical emissivity of radio halos observed at the center of galaxy clusters.

Figure 3 shows the LOFAR image overlaid on the Planck SZ image, where the SZ effect is quantified with the *y*-parameter ($y \propto \int n_e T \, dl$). The radio ridge is located along the filament of gas connecting Abell 0399 and Abell 0401 detected with Planck (*9, 10*). There are hints that the radio emission is not homogeneously distributed, e.g. there are some brighter elongated features that align with the filament direction.

The radio ridge encompassing Abell 0399 – Abell 0401 has unusual properties. Evidence of large-scale shocks in the accretion flows of intergalactic gas has been found in the merging system ZwCl 2341.1+0000 (*16*), whose diffuse radio emission corresponds with an elongated merging cluster of galaxies. The pair of galaxy clusters Abell 0399 – Abell 0401 is at an earlier evolutionary stage, prior to that of ZwCl 2341.1+0000 (*17*). Radio emission associated with the cosmic web joining Abell 0399 and Abell 0401 indicates that some of the merger energy is converted into non-thermal emission, likely through the acceleration of electrons by shock waves and turbulent motions, prior to the collision of the galaxy clusters.

If we assume that the relation between thermal gas density and magnetic field strength found in galaxy clusters (*18*) is also valid in the ridge connecting Abell 0399 – Abell 0401, the thermal gas density of $3\times10^{-4}\,\mathrm{cm}^{-3}$ found in the ridge (*7*) would correspond to a magnetic field strength $B < 1\mu G$. Due to energy lost to synchrotron and inverse Compton radiation, the life-time for relativistic electrons emitting at 140 MHz is $\leq 230$ Myr. The maximum distance these relativistic electrons can travel in their life-time is $\leq 0.1$ Mpc (*2*), more than an order of magnitude smaller than the size of the ridge. There must be a mechanism that re-accelerates and/or injects the electrons in situ, throughout the ridge.

The accretion of matter on large scales along the Abell 0399 – Abell 0401 filament likely injects shock waves and turbulent motion on a wide range of spatial scales. As in the case of radio relics, diffusive shock (Fermi-I) acceleration (DSA) or re-acceleration may power radio emitting electrons and explain this large radio ridge. However, it is difficult to account for such strong emission from shocks, prior to the collision between Abell 0399 and Abell 0401. We explored this scenario using magneto-hydrodynamical simulations with the ENZO code (*19*). Our simulations evolved a pair of merging galaxy clusters at a resolution of 3.95 kpc, with final masses and resultant proximity similar to the Abell 0399 – Abell 0401 complex (*11*). To quantify the expected radio emission, we combined the electrons freshly accelerated at shock waves in the simulation with a radio emission model, which has previously been used to model radio relics (*20 – 22*).

Figure 4 shows the projected gas density in the simulation, after rotating the system to resemble the observed SZ morphology of the Abell 0399 – Abell 0401 complex. Radio emission by freshly accelerated electrons falls $\sim 1000$ times below the sensitivity of our LOFAR observation (Fig. 4A), due to the scarcity of strong shocks in this region (which is filled by a $\sim 5 \times 10^7$ K plasma, heated by gas compression), and to the drop of the assumed electron acceleration efficiency for $M \leq 3$ shocks (*23*). This generates a patchy emission that traces the location of the strongest shocks. In this scenario, the only emission detectable by our LOFAR observation would be associated with a substructure transiting transverse to the line of sight (traced by a weak, $M \sim 3 - 4$ shock similar to region d in Fig. S3). Our LOFAR observation shows emission that is brighter and more broadly distributed across the ridge.

As an alternative model, we tested the additional contribution from a pre-existing population of relativistic electrons, re-accelerated by shocks in the region (Fig. 4B). This is a viable mechanism to illuminate the radio ridge only if the pre-existing electrons that are re-accelerated by shocks with $M \sim 2 - 3$ are accumulated at a Lorentz factor $\gamma \geq 10^3$ and filling most of the ridge volume. This limits the age of these electrons to $< 1$ Gyr due to radiative losses. Circumventing this time constraint would require unidentified volume filling re-acceleration mechanisms in such dilute plasmas.

The non-thermal diffuse emission observed in the Abell 0399 – Abell 0401 system extends far beyond the boundaries of the two radio halos and fills a region in their outskirts which is still dynamically evolving. We interpret this as evidence of intergalactic magnetic fields connecting two galaxy clusters and of spatially distributed particle re-acceleration mechanisms in these regions.

**Acknowledgments:** We thank the referees for their constructive comments which helped improve the presentation of the results. LOFAR, the Low Frequency Array designed and constructed by ASTRON (Netherlands Institute for Radio Astronomy), has facilities in several countries, that are owned by various parties (each with their own funding sources), and that are collectively operated by the International LOFAR Telescope (ILT) foundation under a joint scientific policy. This work is based on observations obtained with Planck (http://www.esa.int/Planck), an ESA science mission with instruments and contributions directly funded by ESA Member States, NASA, and Canada.This research has made use of the NASA/IPAC Extragalactic Database (NED), which is operated by the Jet Propulsion Laboratory, California Institute of Technology, under contract with the National Aeronautics and Space Administration. The Digitized Sky Surveys were produced at the Space Telescope Science Institute under U.S. Government grant NAG W-2166. The images of these surveys are based on photographic data obtained using the Oschin Schmidt Telescope on Palomar Mountain and the UK Schmidt Telescope. H.J. acknowledges the European Commission, Joint Research Center, TP262, Via Fermi, 21027 Ispria (VA), Italy. S.M. acknowledges the LIST Spa - Via Pietrasantina 123, 56122 Pisa, Italy.

**Funding:** A.B. acknowledges support from the ERC Starting Grant DRANOEL GA 714245. The cosmological simulations were performed on Jureca at Jùlich Supercomputing Centre, under computing project HHH42 (P.I. F.Vazza). F.V. acknowledges financial support from the European Union's Horizon 2020 program under the ERC Starting Grant MAG-COW 714196. F.dG., R.J.vW., T.W.S., and H.J.A.R. acknowledge support from the ERC Advanced Investigator programme NewClusters 321271. R.J.vW. acknowledges support from the VIDI research programme with project number 639.042.729, which is financed by the Netherlands Organisation for Scientific Research (NWO). A.M.M.S. acknowledges support from the European Research Council under grant ERC-2012-StG-307215 LODESTONE.

**Author contributions:** All authors meet the journal's authorship criteria. F.G. led and coordinated the project. A.B. worked on the LOFAR data reduction and edited the text. E.O., M.I. and R.P. worked on the LOFAR data reduction. F.V. performed the numerical simulations, worked on the comparison between simulations and radio data and edited the text. M.M. performed the data analysis, edited the text and provided the data at 1.4 GHz. V.V provided the multi-frequencies images (optical, X-ray) and interpreted the radio sources detected in the LOFAR observations. G.G., L.F., F.L., C.F., G.B., R.P., and S.M. interpreted the radio sources detected in the LOFAR observations. C.G. assisted with the numerical simulations. This paper is a collaboration between the group of scientists listed above and a group of the LOFAR Survey Key Project. M.B., G.B., R.C., T.A.E., M.H. provided theoretical expertise on diffuse radio sources in galaxy clusters and are members of the LOFAR Survey Key Project. E.O., A.B., M.I., C.F., R.P., F.dG., C.H., H.J., H.J.A.R., A.S., T.W.S., R.J.vW., and M.W. are all members of the LOFAR Survey Key Project and provided expertise in low frequency data reduction.

**Competing interests:** The authors declare there are no competing interests.

**Data and materials availability:** The observations are available in the LOFAR Long Term Archive (LTA) https://lta.lofar.eu/ under project LC2 005, observed on 15-16 November, 2014. Our simulation has been produced using the public version of Enzo 2.1 https://enzo.readthedocs.io/. The simulation input parameters, are available at

http://doi.org/10.23728/b2share.933d3d24060d4b528c2f6c523ac3844d, and the output data used to produce Fig. 4 and Fig. S7 at
http://doi.org/10.23728/b2share.064065cd568343f1a60135cc49a09e78, both at the EUDAT repository.

**Supplementary Materials:**
Materials and Methods
Figures S1-S7
Table S1
Movies S1-S2
References (*24-54*)

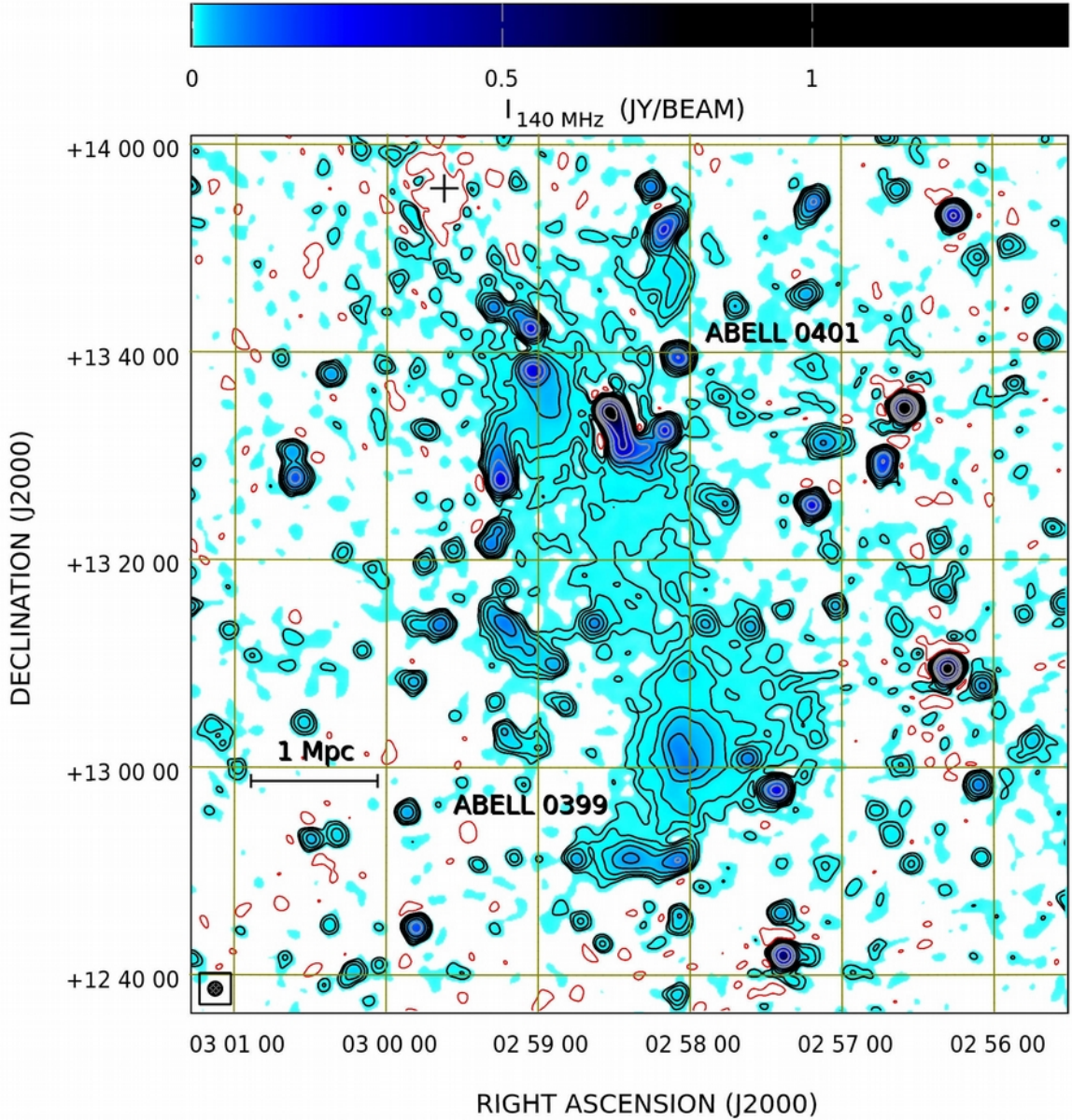

**Fig. 1. LOFAR image of the 1.4°×1.4° region centered on the Abell 0399 - Abell 0401 system.** Color and contours show the radio emission at 140 MHz with a resolution of 80″ and rms sensitivity of 1 mJy beam$^{-1}$. The beam size and shape is indicated by the inset at the bottom left. Contour levels start at 3 mJy beam$^{-1}$ and increase by factors of 2. One negative contour (red) is drawn at −3 mJy beam$^{-1}$. The black cross (right ascension 02h 59m 38s declination +13° 54′ 55″, J2000 equinox) indicates the location of a strong radio source that was removed from the image.

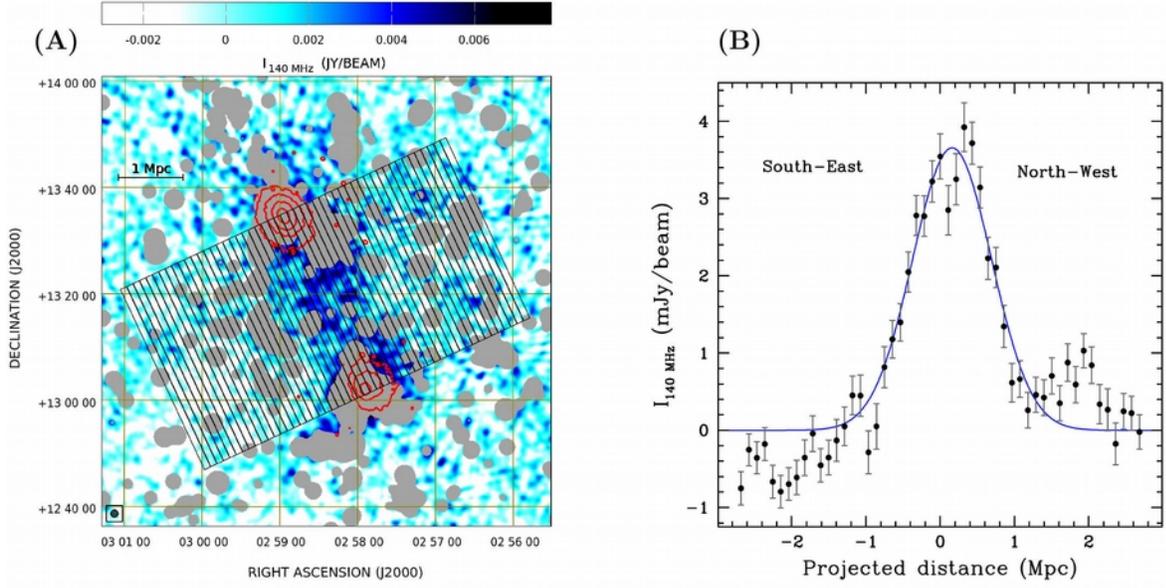

**Fig. 2**. **Brightness profile of the radio ridge**. **(A)**: Strips used to measure the ridge brightness profile. The X-ray emission (*3*) from Abell 0399 and Abell 0401 is overlain in red contours. The color bar represents the LOFAR image in Fig. 1 after masking of sources not related to the radio ridge (grey areas). The width of the strips is 0.108 Mpc (one beam width) and their length is 3 Mpc. The strips are inclined 25° east of the vertical axis and a reference point (at right ascension 02h 58m 26s, declination +13°18 ′17 ″, J2000 equinox) which is located half-way along the line connecting the X-ray positions of Abell 0399 and Abell 0401. **(B)**: Brightness profile of the ridge emission extracted by measuring the average surface brightness in each strip. The error bars are $\sigma/\sqrt{N_{\text{eff}}}$, where $\sigma$ is the image noise rms while $N_{\text{eff}}$ is the number of independent beams in each box, excluding the masked areas. The line represents the best fitting Gaussian model.

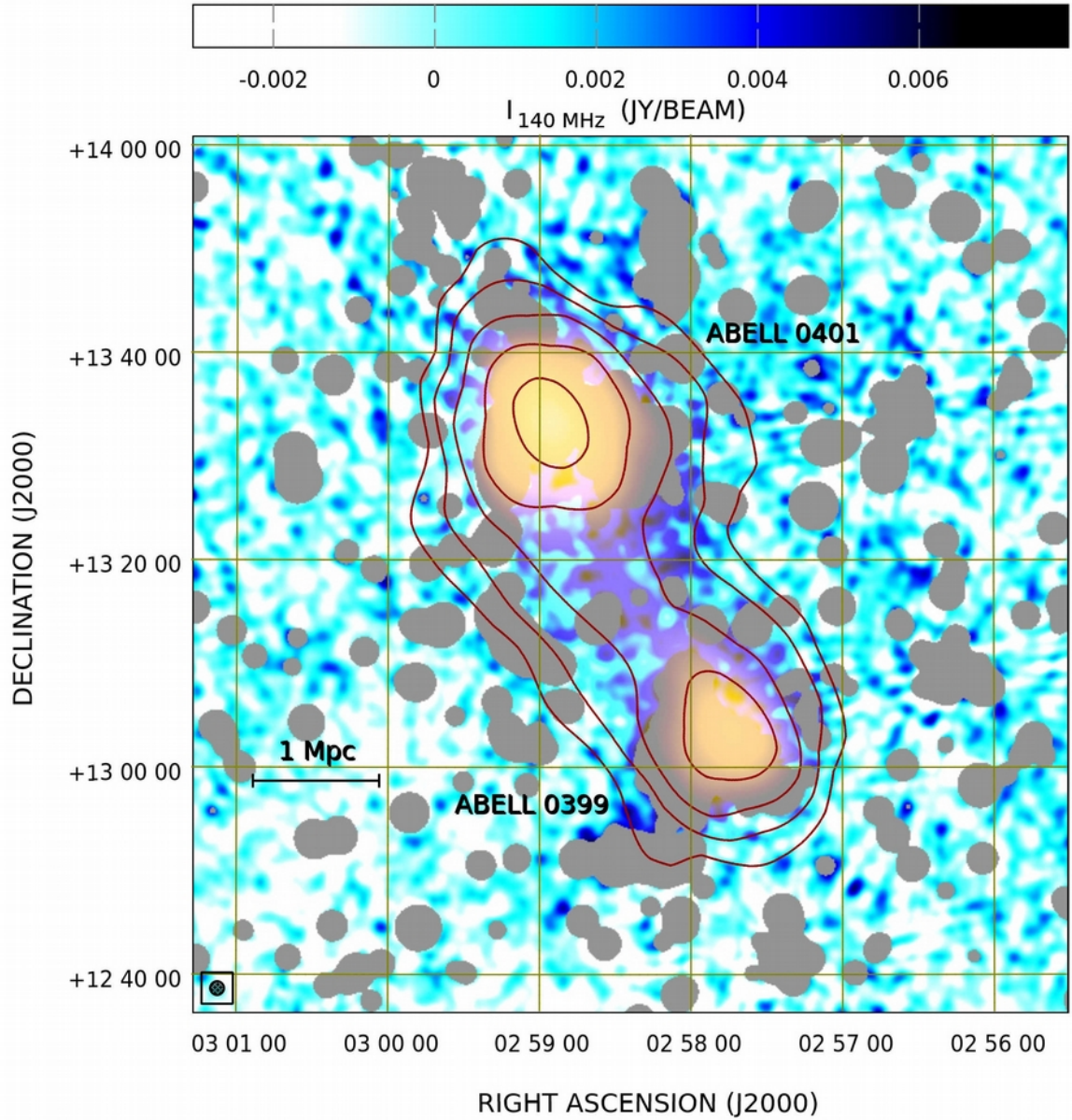

**Fig. 3**. **Composite image showing radio and SZ emission around Abell 0399 – Abell 0401 detected by the Planck satellite**. The same LOFAR image as Fig. 2 is overlain with the Planck *y*-parameter image in yellow tones and brown contours. The Planck data show a bridge of gas between the pair of galaxy clusters Abell 0399 – Abell 0401, in the same location as the LOFAR ridge. Contour levels start at $10^{-5}$ and increase by factors of $\sqrt{2}$.

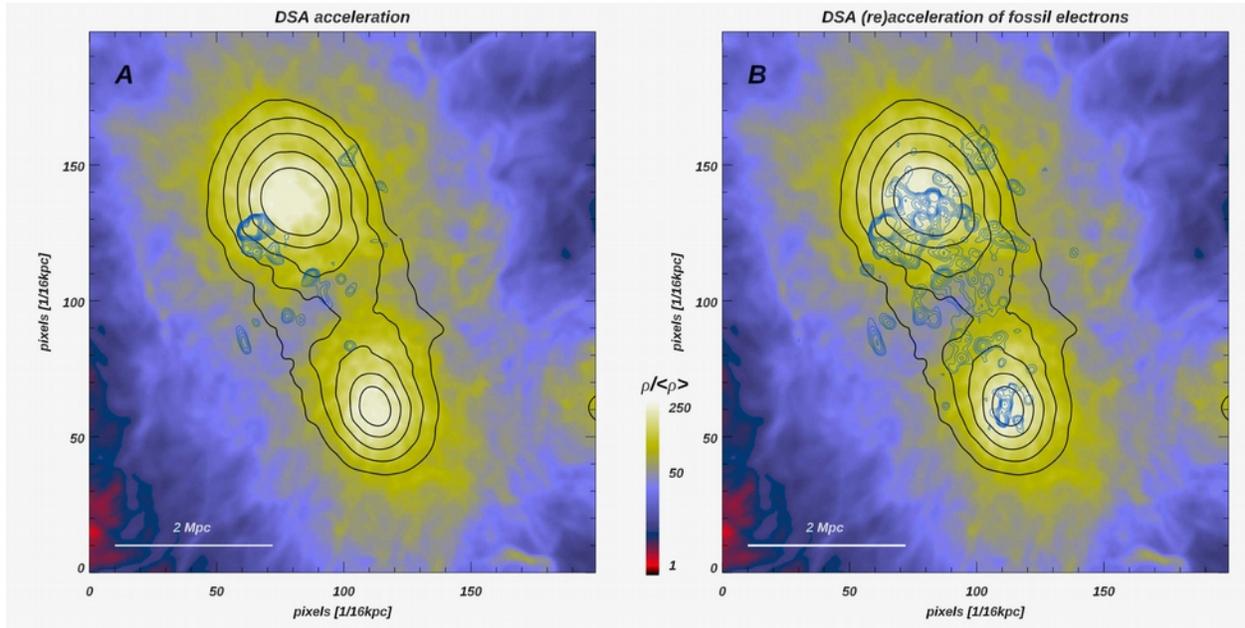

**Fig. 4. Volume rendering of the projected gas density** (colors), **integrated SZ signal** (black contours) **and radio emission** (blue contours) **for our simulated pair of interacting clusters** (*11*), oriented to resemble the Abell 0399 – Abell 0401 pair. Blue contours show the simulated detectable radio emission (≥ 3σ rms of our LOFAR observation) at 140 MHz, for freshly accelerated electrons **(A)** and with an additional contribution of re-accelerated electrons **(B)**. Only the system in (B) resembles the real observation. Movie S1 shows an animated version of this figure.

# Supplementary Materials for

## A radio ridge connecting two galaxy clusters in a filament of the cosmic web

**Authors:** F. Govoni\*, E. Orrù, A. Bonafede, M. Iacobelli, R. Paladino, F. Vazza, M. Murgia, V. Vacca, G. Giovannini, L. Feretti, F. Loi, G. Bernardi, C. Ferrari, R.F. Pizzo, C. Gheller, S. Manti, M. Brüggen, G. Brunetti, R. Cassano, F. de Gasperin, T.A. Enßlin, M. Hoeft, C. Horellou, H. Junklewitz, H.J.A. Röttgering, A.M.M Scaife, T.W. Shimwell, R.J. van Weeren, M. Wise.

\*Correspondence to: federica.govoni@inaf.it

**This PDF file includes:**
Materials and Methods
Figures S1-S7
Table S1

**Materials and Methods**

<u>LOFAR observations</u>

We observed the low frequency radio emission of Abell 0399 (right ascension 02h 57m 56s declination +13° 00′ 59′′, J2000 equinox) and Abell 0401 (right ascension 02h 58m 57s declination +13° 34′ 46′′, J2000 equinox), whose masses (*24*) are ∼$5.7\times10^{14}$ M$_\odot$ and ∼$9.3\times10^{14}$ M$_\odot$, respectively. The redshift (*25*) $z$ of Abell 0399 is 0.071806, while that of Abell 0401 is $z = 0.073664$. At the average distance of the system, 1′ corresponds to 81 kpc in a lambda cold dark matter (ΛCDM) cosmology with: Hubble parameter $h = 0.72$, total matter density $\Omega_M = 0.258$ and cosmological constant $\Omega_\Lambda = 0.742$. The Abell 0399 – Abell 0401 field has been observed (*26*) in the GaLactic and Extragalactic All-sky MWA Survey (GLEAM) at 74-231 MHz. While the extended halos in Abell 0399 and Abell 0401 and other discrete sources were detected by GLEAM, the ridge of radio emission connecting Abell 0399 – Abell 0401 is not.

The Abell 0399 – Abell 0401 system was observed with the LOFAR high band antenna (HBA) stations, on 15-16 November, 2014 under project LC2 005. The total observing time was 10 hours in interleaved mode, switching between scans on the calibrator 3C 48 (5 min) and the target (40 min). We used the full Dutch Array (24 core stations and 14 remote stations), which provides adequate resolution and sensitivity to achive our goals. The HBA dual inner configuration was used. In this configuration, the core stations are split into two HBA sub-fields to gain sensitivity to large-scale emission, and only the inner part of the remote stations is used to have a similar field of view for both the core and remote stations. We used the 8-bit mode that provides a continuous frequency coverage between 110 MHz and 190 MHz. The total bandwidth was split into 440 Sub-bands of 0.18 MHz each, divided into 64 channels. Data were recorded with an integration time of 1s to facilitate removal of radio frequency interference (RFI). To calibrate this dataset we used the prefactor LOFAR pipeline (*27*) for the initial, direction-independent calibration, and Factor for the direction-dependent calibration (*28*). We outline here the main steps of the pipelines. Their details are described elsewhere (*29*).

*Direction independent calibration pipeline:*

The initial calibration consists of RFI-excision, averaging in time to 5s and to 4 channels per Sub-band. Data at frequencies higher than 166 MHz were removed due to excessive interference. The data from the calibrator 3C48 were calibrated against a model (*30*) which sets the flux density scale. We obtained complex gain solutions for the calibrator and solved for a rotation angle per station to take into account differential Faraday Rotation between *XX* and *YY* correlations. The gains derived for the calibrator were used to separate the contribution of the total electron content (TEC) and clock delays per station, and to compute the phase offset between X and Y dipoles. Amplitude gains, station phase offsets, and clock delays were then transferred to the target data. Target data were calibrated in phase against a sky model generated from the TIFR GMRT Sky Survey (TGSS) Alternative Data Release (*31*). A single solution was obtained for every 2 MHz of bandpass. The data were then imaged in bands of 2 MHz at resolution of 30″ and corrected for the effect of the station beam at the phase center. These images have a size that is 2.5 times the FWHM of the station beam. The clean components were subtracted from the data and recorded into a sky model. The source-subtracted data were then re-imaged at lower resolution (~ 1.5′) to remove the contribution of sources in the secondary lobes of the beam and the clean components were subtracted from the data and added to the sky model. These images have a size that is 6.5 times the FWHM of the station beam. The final result of this direction-independent calibration is a dataset almost devoid of sources, divided into bands of 2 MHz each, and a sky model per band. These data and sky models are the inputs to the direction-dependent calibration pipeline.

*Direction dependent calibration pipeline:*

The main goal of this pipeline is to correct for direction-dependent errors across the field of view. These are mainly caused by the ionosphere but also by instrumental effects. For this dataset, the field of view was divided into 50 facets, in addition to a large facet (~ 0.52 degree radius) which contained the target (Abell 0399 and Abell 0401). Every facet contained a bright source (S > 0.3Jy) that was used as a calibrator for that facet. Each facet was processed as follows: data were phase-shifted to the position of the facet calibrator and further averaged in frequency to speed up the calibration process. The model components of the facet calibrator were added back into the visibilities and several cycles of Stokes I phase and TEC self-calibration were performed. Five rounds of complex gain self-calibration were performed. We discarded the last 1.5 h of observations because the TEC solutions showed a very active ionosphere. Once the calibration solutions have converged, these were applied to all the sources within the facet. The whole facet was imaged and the components were subtracted from the visibilities and used to create an updated sky model. The next facet was then processed in the same fashion. In the field of view of our observation, 3 facets were processed before the Abell 0399 – Abell 0401 facet. During the whole process, all baselines shorter than $80\lambda$ – corresponding to angular scales of 43′ – were not used as they were severely affected by RFI. As the angular separation between the clusters is 35′ this baseline cut does not affect the radio ridge.

The images that we present are at a central frequency of 140 MHz and have been obtained at different resolutions. At the resolution of 10″ we reached an rms sensitivity of about 0.3 mJy beam$^{-1}$. In order to reveal extended low surface brightness features, we also produced images at a lower resolution with a higher signal-to-noise (S/N) ratio. In particular, by tapering the longer baselines, we produced a 50″ resolution image with an rms sensitivity of 0.8 mJy beam$^{-1}$ and a 80″ resolution image with an rms sensitivity of 1 mJy beam$^{-1}$, close to the confusion limit expected at this frequency and resolution (*32*). Images were generated using the multi-frequency and multi-scale cleaning method and have been corrected for the differential LOFAR primary beam. The facet containing Abell 0399 and Abell 0401, with width about 1°, includes many other radio sources. Figure S1 shows the diffuse radio ridge encompassing Abell 0399 – Abell 0401 at a resolution of 80″ and the faceting of the field. The faceting has been chosen ad-hoc to have Abell 0399 – Abell 0401 in the same facet. The diffuse emission is well contained within the facet, and restricted to the area in-between the two clusters. The LOFAR flux scale has been cross checked against the TGSS Alternative Data Release. Corrections to the LOFAR absolute flux scale are of a few %, but conservatively, we adopt a 15% uncertainty in agreement with other LOFAR studies (*12*). Images have been generated using the w-stacking clean (WSClean) algorithm (*33, 34*).

The mean emissivity of the radio ridge extrapolated to 1.4 GHz would be $<J>_{1.4GHz} \simeq 4.3 \times 10^{-44}$ erg s$^{-1}$ Hz$^{-1}$cm$^{-3}$ (see main text). Figure S2 shows published distributions of emissivities at 1.4 GHz for candidate large scale filaments (*14*), and radio halos observed at the center of galaxy clusters (*13, 15*). The filament in Abell 0399 – Abell 0401 is located in the weakest tail of the emissivity distribution of the candidates filaments.

LOFAR images at a resolution of 50″ and 10″

Figure S3 shows a composite image of the X-ray emission (*3*) and the radio emission at 50″ resolution, toward Abell 0399 – Abell 0401. Six regions are marked. The emission from the two halos toward the centers of Abell 0399 (region e) and Abell 0401 (region c) is detected at 50″. Both halos are round and regular in shape, although in Abell 0399 an enhancement of the radio intensity is observed, corresponding to a sharp X-ray edge to the east of the cluster core. This was previously observed at 1.4 GHz (*3*), but the LOFAR image reveals that this feature extends much further out into a region of space with faint X-ray emission. The radio halo in Abell 0401 is nestled between several point like, symmetric double, and distorted tailed radio galaxies. Two tailed radio galaxies are located to the south-east (region c) and to the west (region b) of the X-ray core. The radio wake of the south-eastern tail seems to flow directly into the cluster core where it dissolves into the radio halo. Near the edge of Abell 0401, further north-west (region a) from the cluster center, another radio source shows a long tail extending to the south. It may be possible that tailed radio galaxies play a role in supplying relativistic electrons to the cluster environment (*35– 37*), but a clear connection between the tailed radio galaxies and the diffuse emission in Abell 0401 cannot be established with the present data. The two isolated elongated features located to the east of the collision axis, just half way between the two galaxy clusters (region d), and to the south of Abell 0399 (region f ), are not straightforward to classify.

Figure S4 shows zooms of the six regions at a resolution of 50" and 10". The LOFAR contours are overlaid on optical data taken from the red filter of the Digitalized Sky Survey (DSS2 Red). The radio source near the edge of Abell 0401 (region a) is identified with the galaxy 2MASX J02581043+1351519 which has been previously classified as a bent double radio source (*38*). The bright head is followed by a low surface brightness tail extending to the south for about 0.5Mpc. The source to the west of Abell 0401 (region b) shows a bright active core associated with the galaxy PKS 0255+13 at a redshift $z$=0.06466 (*39*), and it is followed by a narrow tail of 0.3Mpc in length extending to the south. A gap is observed between the core and the tail indicating a temporary switch-off of the activity of the radio source (*40*). The tip of the tail makes an abrupt 90° turn to the west and then rapidly fades. The narrow-angle tail (region c) located south-east of Abell 0401 is associated with the galaxy 2MASX J02591487+1327117, at a redshift $z$=0.07753 (*39*). The tail extends to the north and fades into the diffuse halo. A patch of enhanced emission is observed inside the radio halo in Abell 0401, few arcminutes to the north of the central dominant (cD) galaxy. The patch is visible at both 50″ and 10″ and lacks an obvious optical counterpart. The straight feature (region d) located to the east of the collision axis of the Abell 0399 – Abell 0401 system contains two galaxies with known redshifts (*39*). The first (2MASX J02591878+1315467; $z$=0.07276), in the north-east tip is associated with a very weak tailed radio source. The second (2MASX J02591535+1314347; $z$=0.07806), 1.5' to the south-west, has no radio core but has a wake of diffuse radio emission in its trail. We speculate that it could be a switched-off tailed source that is now supplying relativistic electrons. No discrete radio sources are embedded in the radio halo of Abell 0399 (region e) and the brightness enhancement is co-located with the sharp X-ray edge. Finally, the feature (region f) to the south of Abell 0399 contains a very weak tailed radio galaxy (2MASX J02580300+1251138; $z$=0.07481).

We consider the possibility that the radio ridge seen at low resolution is due to the blend of faint discrete sources undetected in the LOFAR image at 10″ resolution but that add up in the LOFAR image at 80″ resolution. We consider the 214 galaxies (*39*) detected in the field of view of Abell 0399 – Abell 0401 with redshifts in the range from $z$=0.063 to $z$=0.083. Of these, 34 are located in the ridge area, as shown in Fig. S5A. To increase the S/N ratio, we stack the emission of these galaxies using sub images of 48″ × 48″ in size. Figures S5B shows the stacking of the optical emission for the 34 galaxies located in the ridge area. Figures S5C shows the stacking of the radio emission at 10″ resolution for the same galaxies. The noise level of the radio image is then reduced to 0.3/ √34 = 0.05 mJy beam$^{-1}$. From the stack of the LOFAR radio emission we find that, on average, the galaxies in the ridge area have a flux density of 0.47 mJy. In order to reproduce the average ridge intensity of 2.75 mJy beam$^{-1}$ measured at 80" resolution, we would need about 5.8 galaxies per beam. Since the ridge between the two clusters (inside an area of 2 × 1.3 Mpc$^2$) is detected in 138 beams, the galaxies needed in the area would be an implausible total number of 800.

Spectral analysis

We derived the spectral index distribution of the radio sources in the field of Abell 0399 – Abell 0401 by comparing the 140 MHz LOFAR observation at 50″ with a previously-published Very Large Array (VLA) image at 1.4 GHz (*3*). We smoothed the VLA image to 50″ in order to match the resolution of the LOFAR image and computed the spectral index and its uncertainty for all pixels whose brightness is above a 5σ threshold in both images. The rms uncertainty in the spectral index is generally less than 0.12, with systematic uncertainties of 15% and 5% at 140 MHz and 1.4 GHz, to account for the uncertainty on the flux density scale calibration. In Fig. S6A-B we show the distribution of the spectral index and its rms uncertainty for a region around Abell 0401. In Fig. S6C-D we show the corresponding region surrounding Abell 0399. The lowest isophote in the LOFAR image is not detected in the VLA image if the radio spectrum is steeper than α > 1.3. As a result, these faint steep-spectrum regions are not represented in our spectral index image. The three tailed radio sources surrounding Abell 0401 show the typical spectral gradients usually observed for these radio sources, with an α that increases from 0.5, close to the active galactic nucleus, up to ~2.0 at the end of the tail. If we consider the regions detected above a 5σ threshold at both 140 MHz and 1.4 GHz, the radio halo in Abell 0401 has an average spectral index <α> = 1.4 ± 0.1. At the position of the X-ray peak the halo spectral index is flatter, α = 1.2 ± 0.1, but the patch of enhanced brightness north east to the galaxy cluster core has a much steeper spectrum with α=2.1 ± 0.1. The halo in Abell 0399 has the same average spectral index of <α> = 1.4 ± 0.1. Close to the cluster X-ray peak the radio spectrum is comparatively flatter, with α = 1.0 ± 0.1, but the eastern part of the halo close to the sharp X-ray edge has a steeper spectrum with α = 1.8 ± 0.1. The spectral index of the straight feature located in the filament connecting Abell 0399 – Abell 401 (region d in Fig. S3) is determined only close to the north-east tip, i.e. in the region occupied by the two faint tailed sources shown in Fig. S4. Here the spectral index shows a typical gradient for the northern active tailed source and a slightly steeper spectrum for the switched-off tail in the south. The rest of the source is not detected at 1.4 GHz, and we can only place a lower limit of α > 1.7 to the spectral index. The spectral index of the structure to the south of Abell 0399 (region f in Fig. S3) presents the typical gradient for a tailed source with α≃0.5 close to the active galactic nucleus in the right end of the source.

To reduce the bias of the spectral index image toward flat spectrum regions, we also calculate the global spectral indexes by integrating the radio intensity within the area delimited by the 140 MHz 5σ isophote, thus including also the steep spectrum regions not detected at 1.4 GHz in the image shown in Fig. S6. For the halo in Abell 0401 we measure an unbiased global spectral index of <α> = 1.63 ± 0.07 while for the halo in Abell 0399 we obtain <α> = 1.60 ± 0.07. Using the same approach for the sources in regions d and f we measure <α> =1.23 ± 0.07 and <α> = 0.96 ± 0.07, respectively. The global flux density measurements, the spectral indexes, the area, and the largest angular size (LAS) for these sources are listed in Table S1.

Current data do not permit us to classify the newly detected straight features (region d and region f) unambiguously. The data at high resolution and the spectral index images confirm the presence of head tails and a switched-off tailed radio galaxy in these regions. Therefore, they might be faint radio galaxies whose tailed morphology were not detected in previous observations. On the other hand, we do not exclude the possibility that they might be relic sources similar to the case of PLCK G287.0+32.9 (*41*) in which the radio relic appears to be connected to a radio galaxy. Relativistic electrons in relics are likely to be produced by Diffusive Shock (Fermi-I) Acceleration (DSA) or re-acceleration at the merger shocks (*23, 42–45*). The morphology and the size (> 700 kpc) of these sources is in agreement with the relic classification. In support of this classification, the Suzaku satellite (*7*) detected a temperature discontinuity in region d possibly associated with a weak shock wave caused by an accretion flow onto the filament connecting Abell 0399 – Abell 0401. On the other hand, in radio relics we expect a gradient in spectral index transverse to the major axis, which is not seen. The picture is further complicated by the fact that with the current data faint tailed radio galaxies and relics are not easily distinguishable from other types of cluster radio sources like radio phoenices which are believed to be the result of the re-energization via adiabatic compression, triggered by shocks, of fossil plasma from switched-off radio galaxies (*35, 46–48*).

Numerical Simulations

We simulated a pair of colliding clusters with the cosmological grid code ENZO (*19*), using the Dedner method (*49*) to solve magneto-hydrodynamics (MHD) on a comoving grid, and employing adaptive mesh refinement (AMR) to selectively increase the spatial and force resolution down to 3.95 kpc/cell in most of the virial volumes of our clusters (*50)*. The two clusters form in a total volume of (260 Mpc)$^3$ (comoving), which is simulated with two nested levels containing 256$^3$ cells and dark matter particles, for an initial spatial resolution of 507 kpc/cell in the innermost (130 Mpc)$^3$ region centered onto the clusters formation region. We allowed the selective increase of spatial/force resolution during run-time with AMR, whenever the gas density in a cell is ≥ 1% higher than its surroundings, limited to the innermost ∼ 25 Mpc$^3$ volume of the box. Most of the virial volume of the two clusters is covered by cells with a resolution from 3.95 to 7.9 kpc at $z = 0$. Our simulation includes the effects of cosmic expansion, gravity and MHD, while for simplicity we neglect the effect of radiative gas cooling and feedback. The magnetic field is assumed to have a primordial origin, and is initialized at the beginning of the run ($z = 30$) as a uniform field along the three coordinate axes of the simulation, with strength $10^{-4}$ μG (comoving) seed field injected at the initial redshift of the simulation. The resolution probed in this run is high enough to produce a small-scale dynamo, which increases the magnetisation in the cluster cores to the ∼ μG level (*50,51)*, and to several ∼ 0.1μG in the contact region of the two clusters.

Shock waves are identified in post-processing using a shock finder that tags 3-dimensional velocity jumps and compute the Mach number based on standard Rankine-Hugoniot jump conditions (*52*).

To predict the synchrotron emission from accelerated electrons, we assume an exponential cutoff in the energy distribution of electrons, determined by the balance of the acceleration rate from DSA and of the synchrotron and inverse Compton cooling rate (*23*). In the post-shock region, DSA is assumed to generate supra-thermal electrons that follow a power-law in momentum. The total radio emission at 140 MHz is computed from the contributions of all electrons accelerated in the post-shock. In the standard DSA scenario in which electrons that emit synchrotron radiation are accelerated out of the thermal pool, the acceleration efficiency as a function of Mach number and gas temperature, $\xi(M,T)$ is $\xi \approx 10^{-7}$ for M = 2.5 shocks with an upstream temperature of T = $10^7$ K, and $\xi \approx 5 \cdot 10^{-4}$ only for M ≥ 10 shocks in the same environment (*23*). We estimated the additional contribution from re-accelerated (fossil) electrons by upscaling the $\xi(M,T)$ function, based on the expected re-acceleration efficiency of cosmic ray protons (*53*) and assuming a uniform $10^{-3}$ energy ratio between relativistic electron and proton in the shock region. As discussed in the main paper, this approximation is valid only for a young enough population of fossil electrons to have a power-law spectrum up to $\gamma \sim 10^3$. In this case, the effective re-acceleration efficiency of electrons increases to $\xi \approx 10^{-5}$, i.e. the contribution from re-accelerated electrons outshines that of freshly injected electrons (*54*). Figure S7 shows the outcome of the re-acceleration scenario and the projected distribution of magnetic fields in our simulation. Movie S1 and Movie S2 show the 3-d simulated pair of clusters. The mock LOFAR observation at 140 MHz was produced by convolving the simulated emission with a beam of ∼80″ and a Gaussian noise of ∼1 mJy beam$^{-1}$, and by locating the source at the redshift corresponding to the real Abell 0399 – Abell 0401 complex.

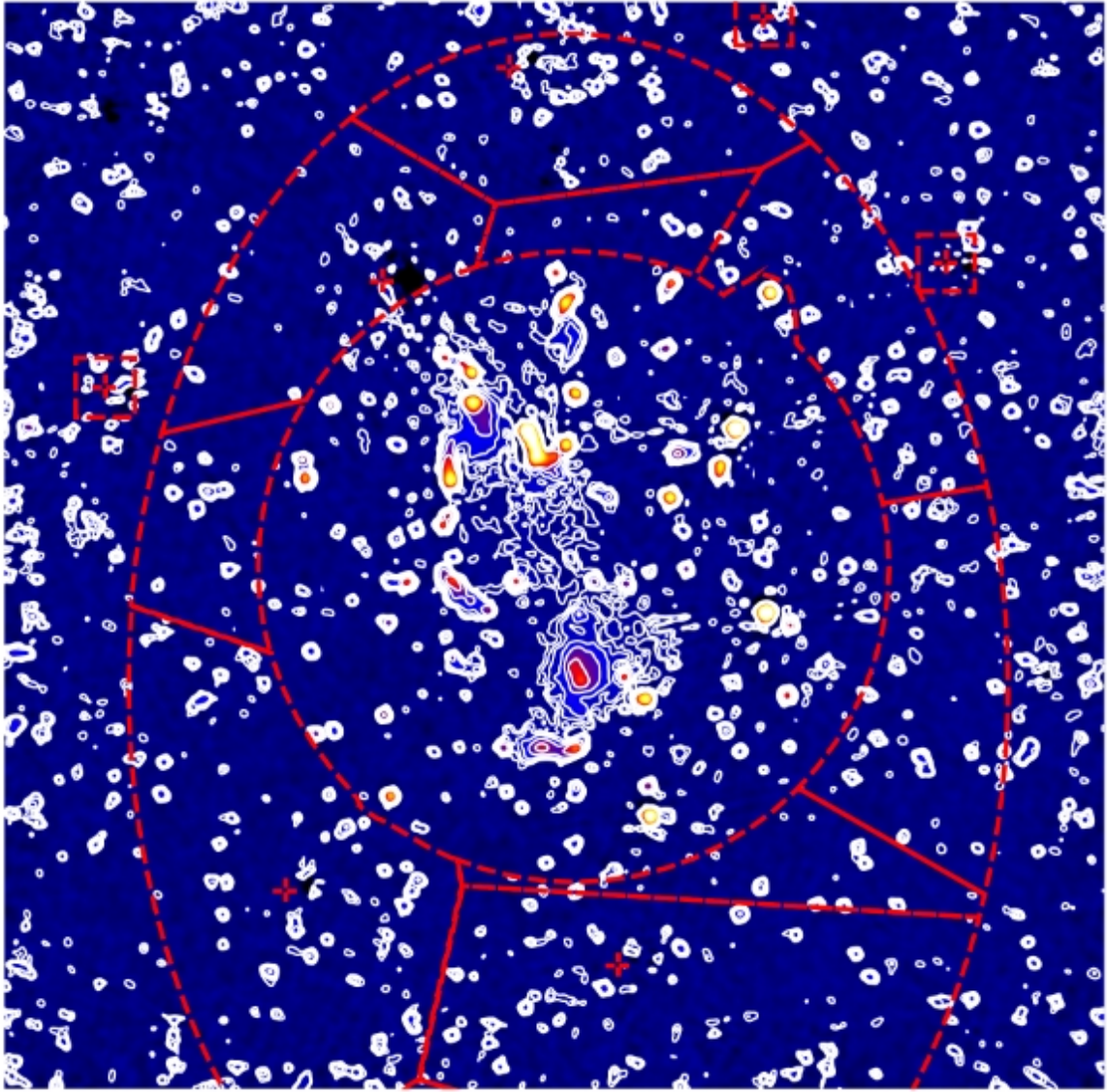

**Fig. S1. LOFAR image centered on the Abell 0399 – Abell 0401 system.** Color and contours show the radio emission at 140 MHz with a resolution of 80″ and rms sensitivity of 1 mJy beam$^{-1}$. Contour levels start at 3 mJy beam$^{-1}$ and increase by factors of 2. The faceting of the field is shown with red lines.

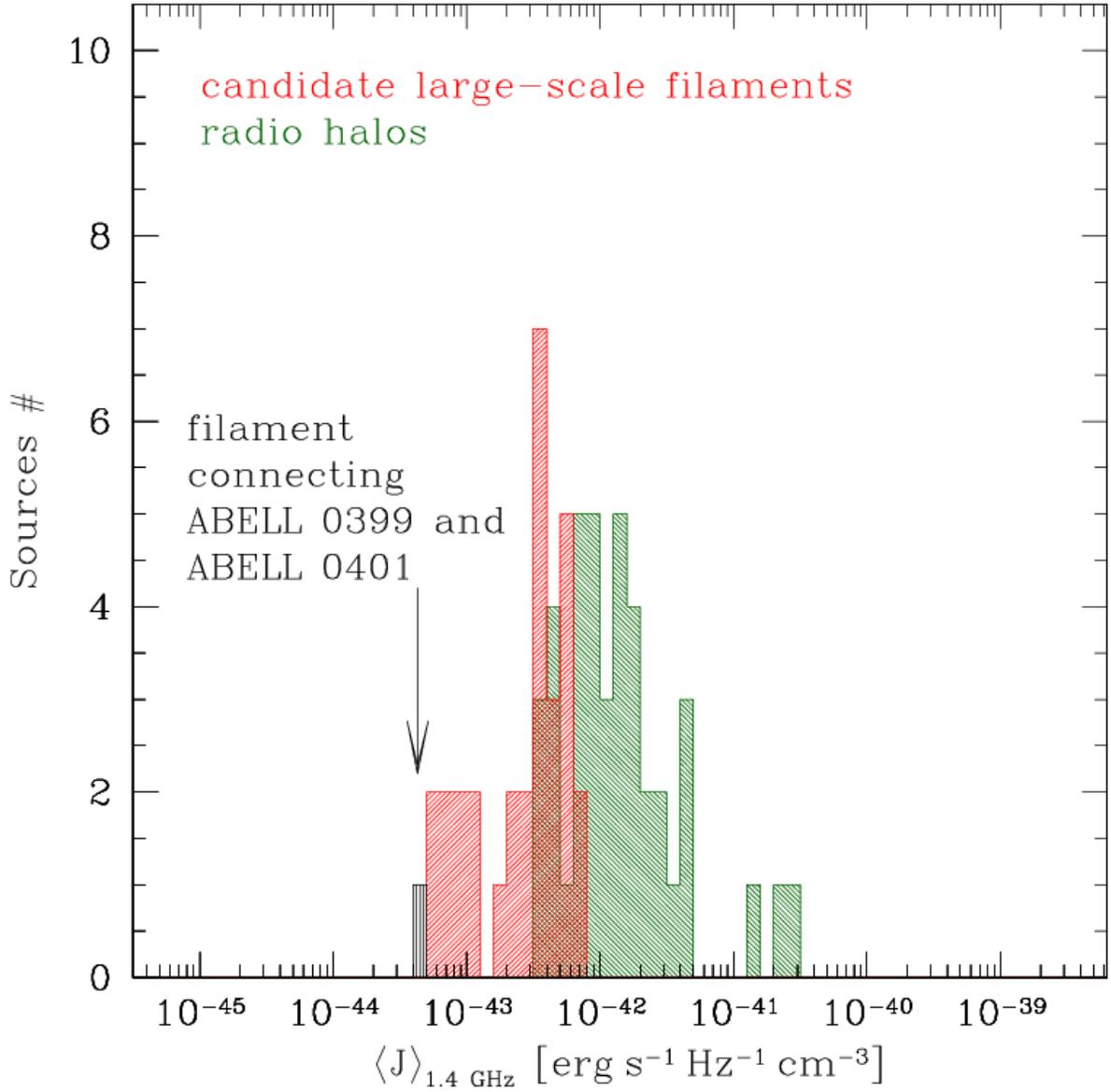

**Fig. S2. Histogram in logarithmic scale of the mean emissivity** of the candidate large-scale filaments in red (*14*) compared to the cluster radio halos in green (*13,15*). The emissivity of the filament connecting the galaxy clusters Abell 0399 and Abell 0401 extrapolated at 1.4 GHz is shown in black (see also the vertical arrow).

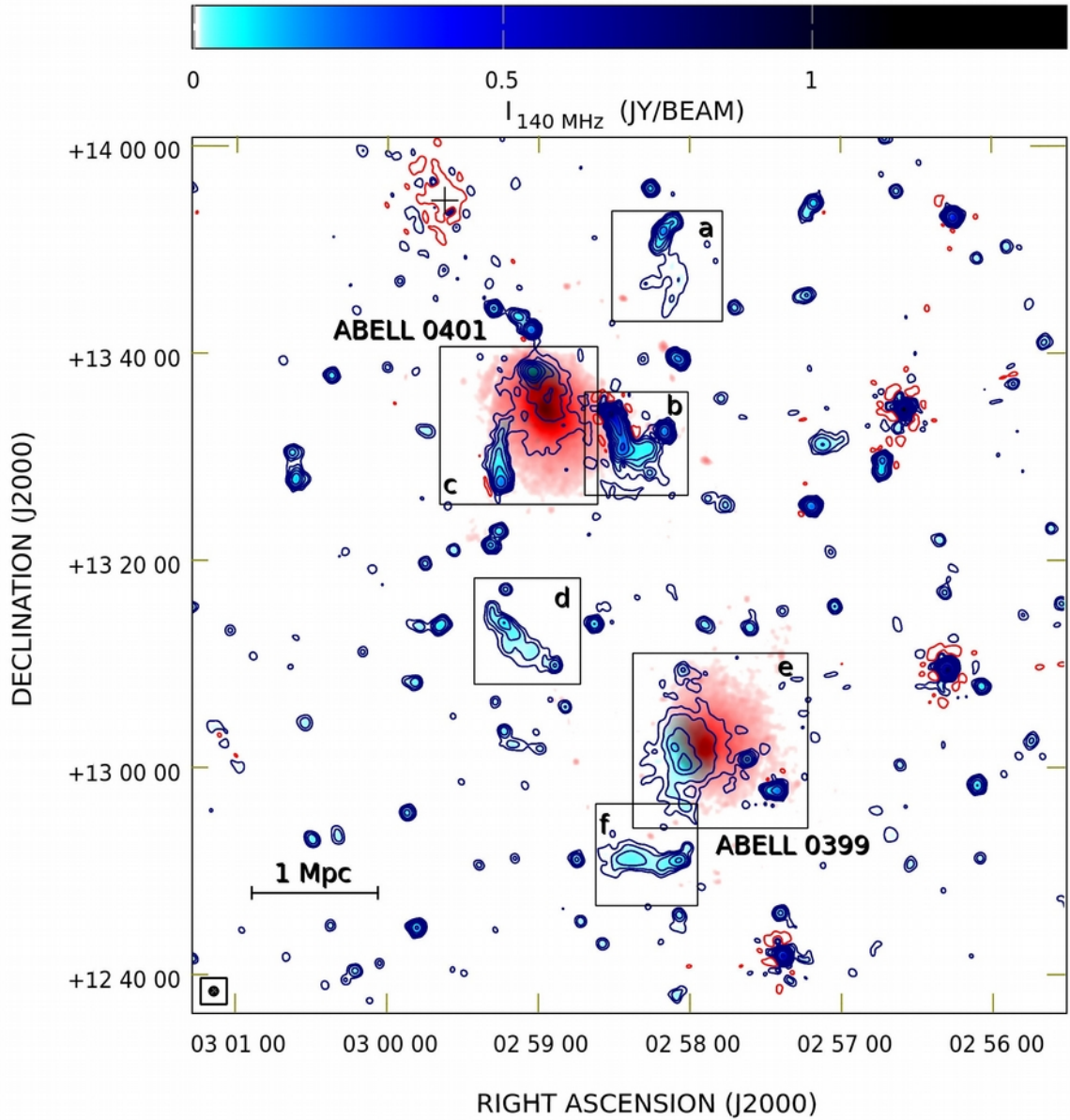

**Fig. S3. Composite image showing the low-frequency radio emission and the X-ray emission toward Abell 0399 – Abell 0401**. In red tones the X-ray image, obtained (*3*) with the XMM-Newton satellite, in the 0.2–12 keV band. The X-ray image has been smoothed with a Gaussian kernel of σ=12″. Blue tones and contours represent the radio emission as obtained with LOFAR at 140 MHz. The radio image has a resolution of 50″. The beam size and shape is shown by the inset in the bottom left. Contour levels start at 4 mJy beam$^{-1}$ and increase by factors of 2. One negative contour (red) is drawn at –4 mJy beam$^{-1}$. The rms sensitivity of the radio image is 0.8 mJy beam$^{-1}$. Regions described in the text are labeled and marked with boxes. The black cross (right ascension 02h 59m 38s declination +13° 54′ 55″, J2000 equinox) indicates a strong radio source removed from the image.

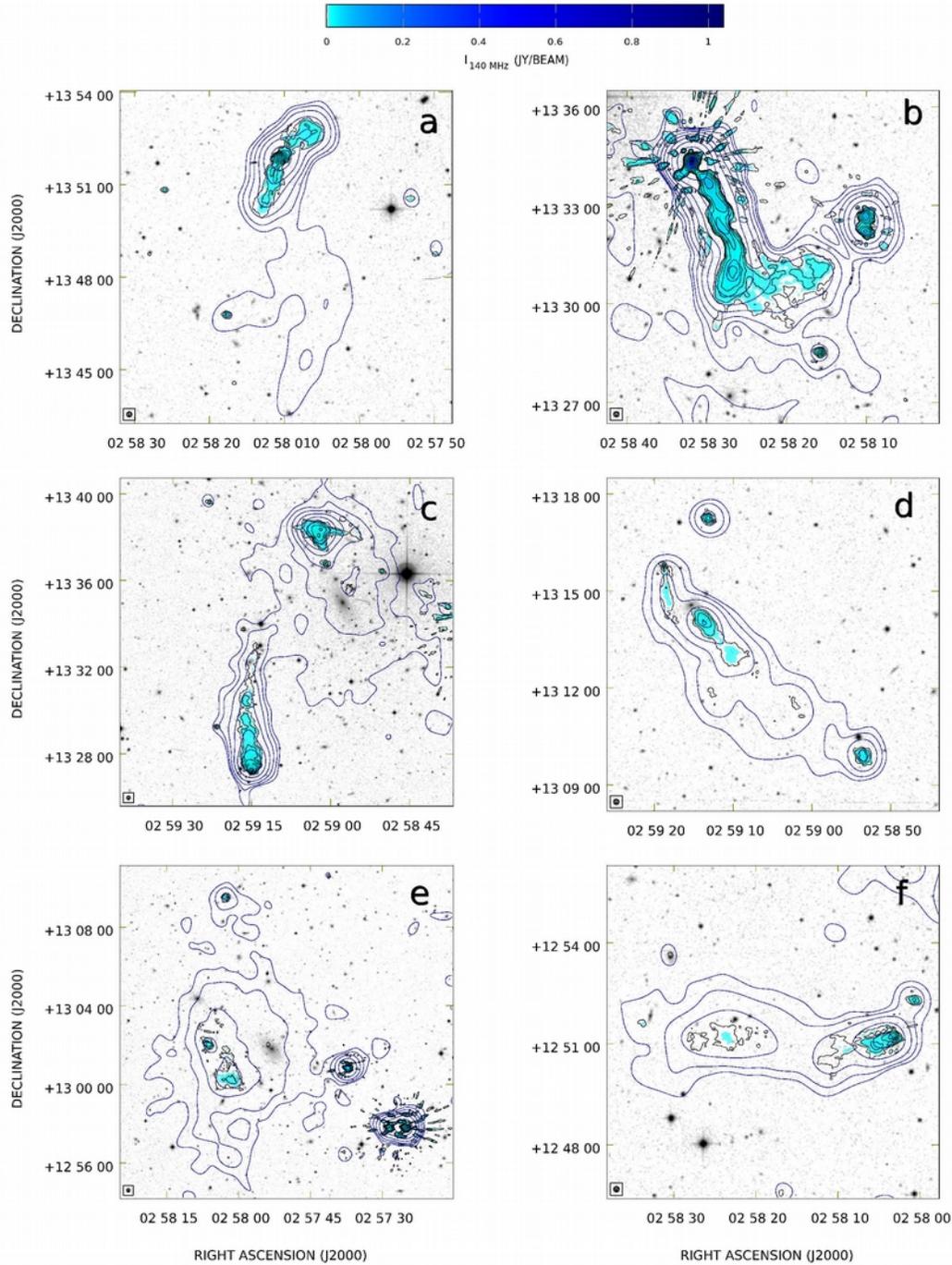

**Fig. S4. Zoomed versions of the regions marked in Fig. S1.** The background grayscale image shows optical data from the Digital Sky Survey (DSS2 Red). Blue contours are the same as shown in Fig. S1. In blue tones and black contours, we show the radio emission detected at 10″ of resolution with LOFAR at 140 MHz. The first contour level is drawn at 1.5 mJy beam$^{-1}$, and the rest are spaced by factors of 2. The sensitivity is 0.3 mJy beam$^{-1}$. This image is affected by some artifacts of bright sources in the field e.g. bottom right of region e.

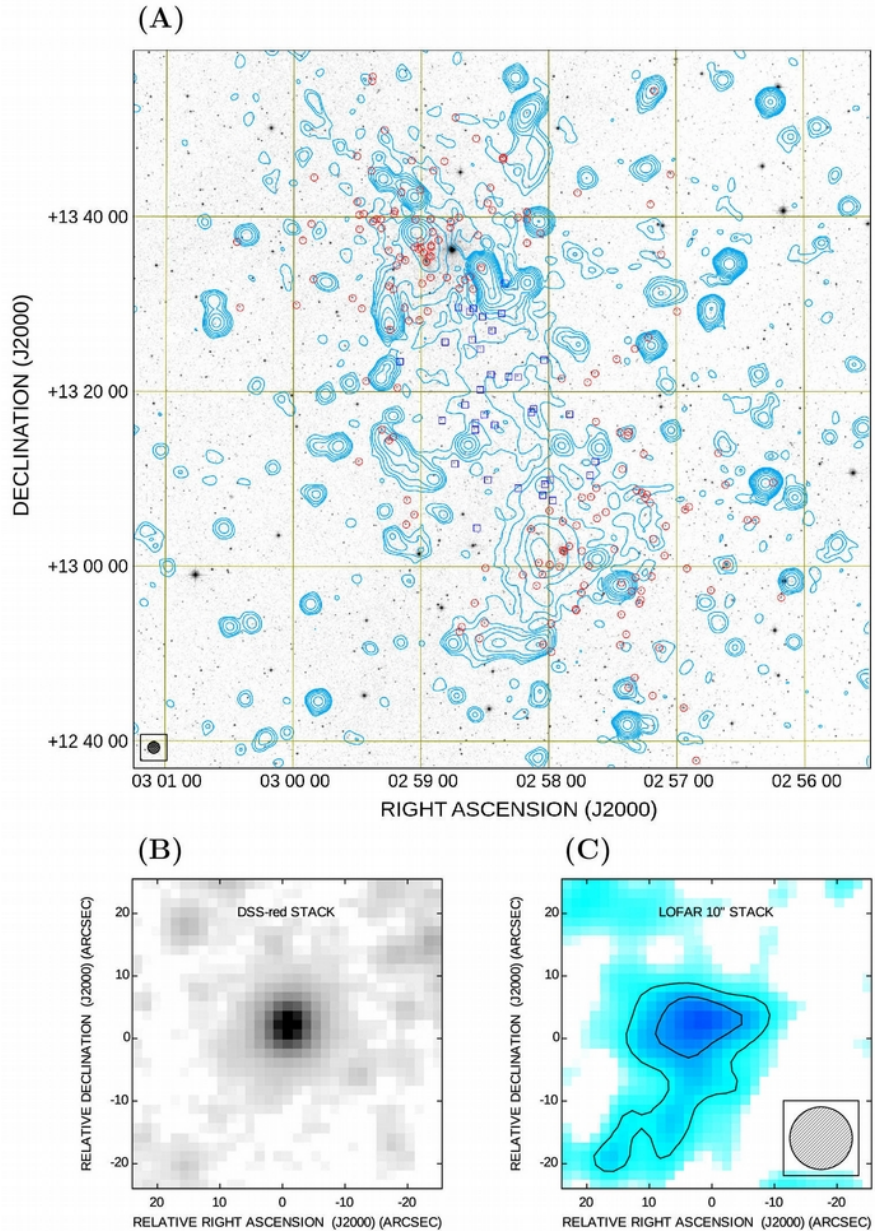

**Fig. S5. Galaxies in the Abell 0399 – Abell 0401 system with redshift in the range from z=0.063 to z=0.083. (A)**: Contours show the LOFAR emission at 140 MHz with a resolution of 80″. The beam size and shape is shown by the inset in the bottom left. Contour levels start at 3 mJy beam$^{-1}$ and increase by factors of 2. The radio emission is overlaid on the optical DSS2 image. Red circles and blue squares refer respectively to the galaxies outside and inside the 2 × 1.3 Mpc$^2$ ridge area between the two clusters. **(B)**: Stacking of the optical emission for the 34 galaxies located in the ridge area. **(C)**: Stacking of the radio emission for the 34 galaxies located in the ridge area. The beam shown in the bottom-right corner is 10″. The first contour level is at 0.25 mJy beam$^{-1}$ (i.e. the 5σ level in the stacked image) and the rest are spaced by factors of √2.

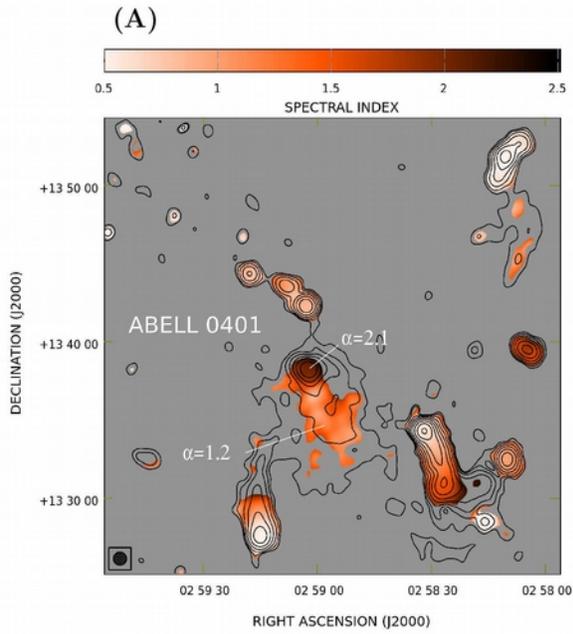 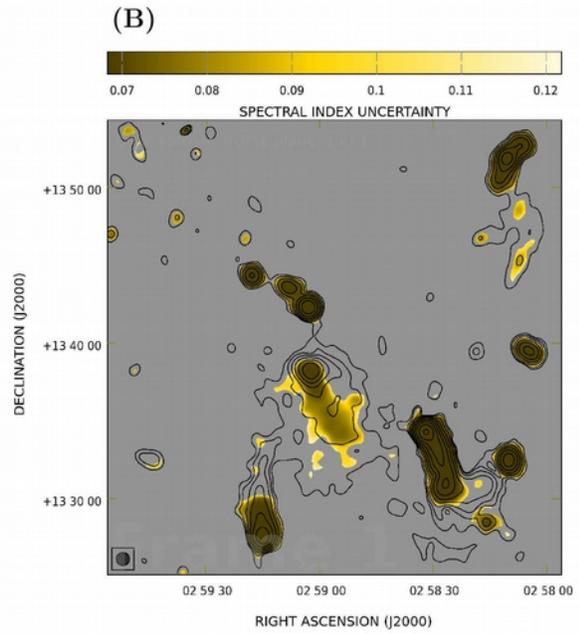
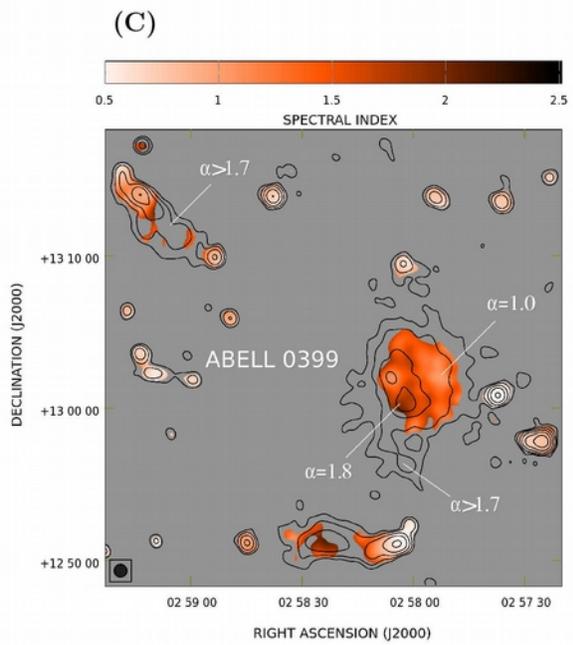 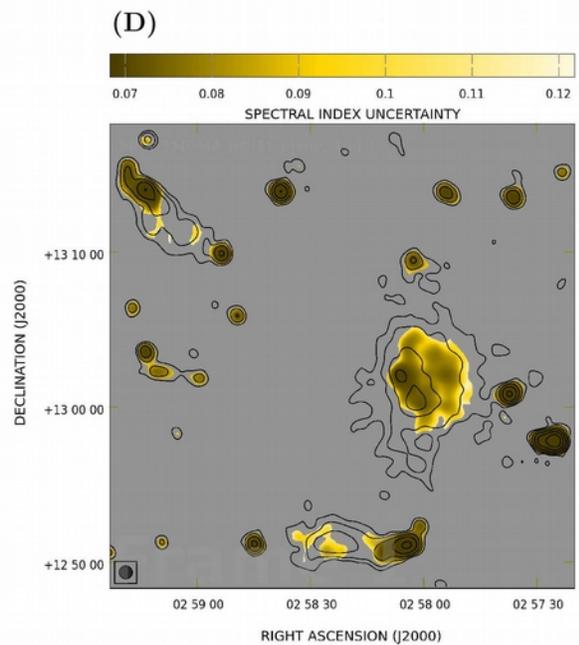

**Fig. S6. (A):** S**pectral index image calculated between 140 MHz and 1.4 GHz** for a region around Abell 0401. **(B):** Spectral index rms uncertainty for a region around Abell 0401. **(C) - (D):** Show the same for the region surrounding A399. Contours refer to the LOFAR 50″ image. Levels start from 4 mJy beam$^{-1}$ and increase by factors of 2.

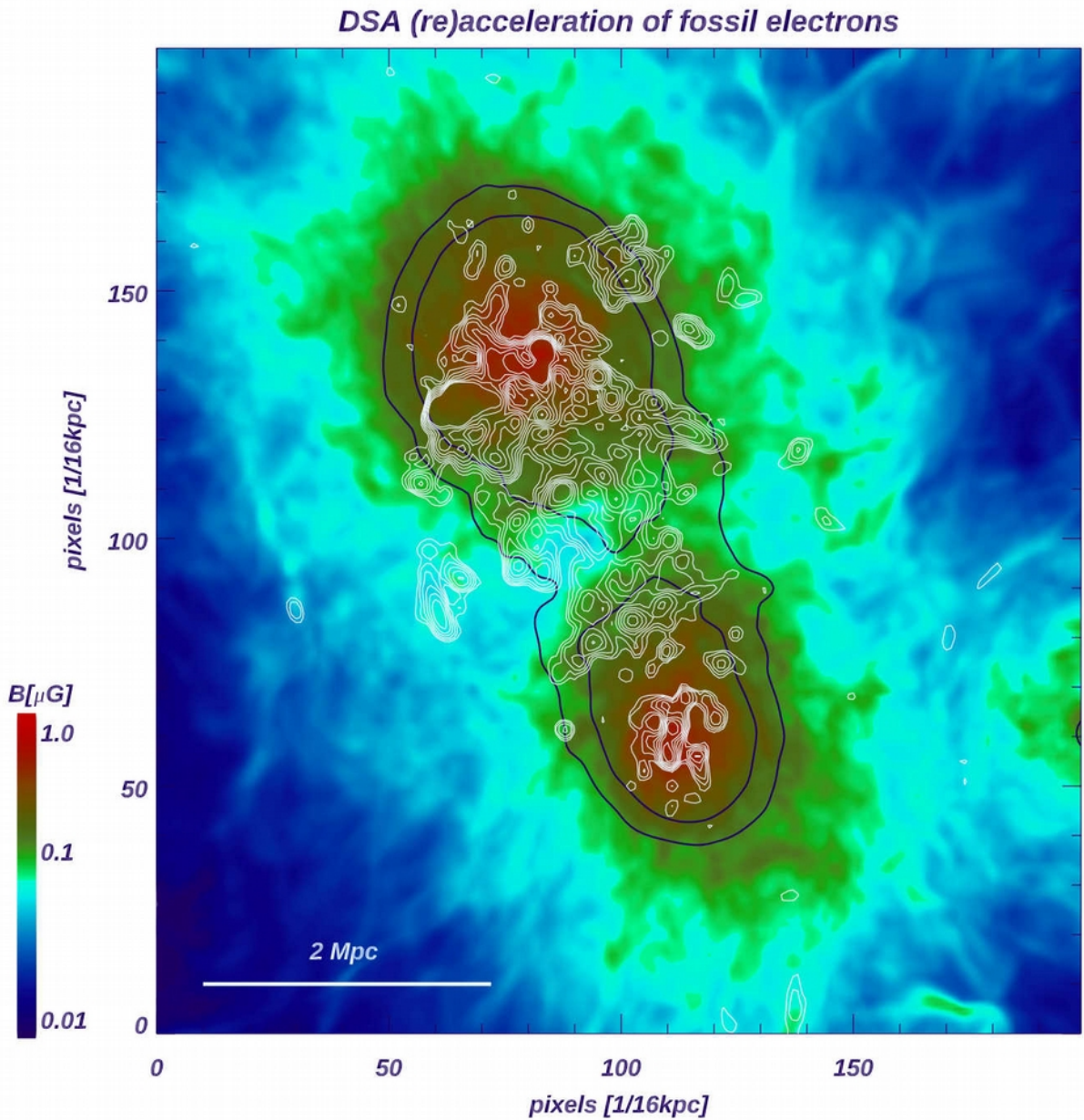

**Fig. S7. Volume rendering of the projected magnetic field strength** (colors), **of the integrated SZ signal** (black contours) **and of the radio emission** (white contours, same as Fig.4B) **for our simulated pair of interacting clusters.** Movie S2 shows an animated version of this figure.

Table S1. **Flux density measurements at 140 MHz and 1.4 GHz** calculated by integrating the radio intensity in the area delimited by the 140 MHz 5σ isophote. The quoted uncertainties refer to the 1σ rms statistical error. Systematic uncertainties of 15% and 5% (indicated in parentheses), take into account the uncertainty on the calibration of the LOFAR and VLA flux scale respectively. These are included in the evaluation of the total uncertainty on the quoted spectral indexes.

| Source | $S_{140MHz}$ (mJy) | $S_{1.4GHz}$ (mJy) | $L_{140MHz}$ (W/Hz) | $\langle\alpha\rangle$ | Area (arcmin$^2$) | LAS (arcmin) |
|---|---|---|---|---|---|---|
| A401 | 826 ±2.5(±124) | 19.3 ± 0.6(±1.0) | $1.0 \times 10^{25}$ | 1.63 ± 0.07 | 46.9 | 10 |
| A399 | 807 ±2.7(±122) | 20.4 ± 0.6(±1.0) | $9.8 \times 10^{24}$ | 1.60 ± 0.07 | 60.4 | 12 |
| Region d (relic candidate) | 339 ±2.0(±51) | 19.9 ± 0.5(±1.0) | $4.1 \times 10^{24}$ | 1.23± 0.07 | 21.5 | 9.5 |
| Region f (relic candidate) | 384 ±2.0(±58) | 42.1 ± 0.5(±2.1) | $4.7 \times 10^{24}$ | 0.96 ± 0.07 | 21.0 | 8.8 |

**Movie S1. A merging pair of galaxy clusters (at high resolution).** Projected gas temperature (left) and average magnetic field strength along the line of sight (right) for a simulated pair of colliding clusters, before their first encounter. The simulation has been produced by using the ENZO code (*19*).

**Movie S2. A merging pair of galaxy clusters (at high resolution).** Projected Y-compton parameters (colors) and projected radio emission (contours) detectable at 140 MHz assuming a ~ mJy beam$^{-1}$ sensitivity with LOFAR-HBA. The left panel shows a model in which the radio emission is only produced by electrons which are directly accelerated from the thermal pool via diffusive shock acceleration, while the right panel gives the radio emission after including the additional contribution by "fossil" relativistic electrons, which get reaccelerated by shocks. The simulation has been produced by using the ENZO code (*19*).